\documentclass[12pt]{iopart}

\expandafter\let\csname equation*\endcsname\relax
\expandafter\let\csname endequation*\endcsname\relax

\usepackage[T1]{fontenc} 

\usepackage{lmodern}       
\usepackage{microtype}    
\usepackage{booktabs}
\usepackage{subfig}    
\usepackage{amssymb,amsfonts}
\usepackage{graphicx}
\usepackage{bm}
\usepackage{hyperref}
\usepackage[dvipsnames]{xcolor}

\usepackage{amsmath}
\usepackage{comment}
\usepackage{dsfont}
\usepackage{mathtools}

\usepackage{cite}

\DeclareUnicodeCharacter{02C6}{^}

\usepackage{braket}

\usepackage{cleveref}      
\begin{document}

\title[Tunneling in QC: WKB vs SWKB]{Tunnelling in Quantum Cosmology: WKB vs SWKB}
\author{Duarte Guimar\~aes$^{1,2,*}$, Jo\~ao Marto$^{1,2}$ and Paulo Vargas Moniz$^{1,2}$}

\makeatletter
\def\@fnsymbol#1{\ifcase#1\or *\or \dagger\or \ddagger\or \S\or \P\or \|\or **\or \dagger\dagger\or \ddagger\ddagger \else\@ctrerr\fi}
\makeatother

\footnotetext[1]{Corresponding author: Duarte Guimarães; Email: duarte.guimaraes@ubi.pt}

\address{$^1$ Departamento de F\'{\i}sica, Universidade da Beira Interior, Rua Marqu\^{e}s D'\'Avila e Bolama, 6200-001 Covilh\~a, Portugal}
\address{$^2$ Centro de Matem\'atica e Aplica\c{c}\~oes da Universidade da Beira Interior, Rua Marqu\^{e}s D'\'Avila e Bolama, 6200-001 Covilh\~a, Portugal}

\vspace{10pt}
\begin{indented}
\item[]\today
\end{indented}

\begin{abstract} This study analyses the WKB and Supersymmetric WKB (SWKB) methods within a closed Friedmann-Robertson-Walker (FRW) minisuperspace model to determine whether these approaches yield different results in modelling tunnelling phenomena in quantum cosmology.
The transition from a dust-dominated to a dark-energy-dominated epoch driven by a generalised Chaplygin gas is examined. Incorporating supersymmetric quantum mechanics requires modifying the initial phenomenological potential, which in turn influences late-time cosmological evolution. Specifically, after the dust-dominated phase, the universe enters a dark-energy-dominated dynamical regime.
This regime decays over time and eventually returns to another dust-dominated period.
Analytic approximations for the superpotential, including power series and the Picard approximation, yield closed-form SWKB tunnelling expressions. These expressions facilitate the calculation of transmission probabilities as functions of the Chaplygin parameters $A$, $B$, and $\alpha$.
The SWKB and WKB methods can both be employed for a concrete range of the scale factor, $a$ and the Chaplygin gas parameters, but quantitatively they provide different tunnelling probabilities. In more detail, they diverge for suppressed barriers: standard WKB systematically underestimates the tunnelling probability, while SWKB mathematically avoids the turning-point singularity but physically overcompensates due to the lack of shape-invariance. A backward numerical integration method is introduced as a numerical benchmark, confirming that the true transmission probabilities lie strictly between the two semiclassical approximations ($T_{WKB} < T_\text{num} < T_{SWKB}$). Crucially, by confirming that the actual tunnelling probabilities are higher than estimated by standard WKB, these results demonstrate that quantum transitions into cosmological models with transient late-time acceleration are probabilistically more viable.
Therefore, SWKB provides a valuable complementary approach for studying barrier transmission in quantum cosmology. It is suggested that a functional form for the decaying dark energy density should be derived. This would enable a test of the SQM modified potential and of its implications with current observational data.
\end{abstract}

\vspace{2pc}

\submitto{\CQG}

\section{Introduction}

Supersymmetric Quantum Mechanics (SQM) was first introduced by Edward Witten \cite{WITTEN1981513} as a framework to study supersymmetry breaking and related phenomena in quantum field theory.
SQM quickly attracted broad interest \cite{COOPER1983262} and remains an active research area, particularly as a framework to study
quantum mechanical problems \cite{Khare:2004kn,gangopadhyaya2024}. Although not formulated as SQM, Erwin Schr\"odinger was the first (1940) to obtain the
eigenvalues and eigenfunctions of the hydrogen atom by factorising the Hamiltonian into the product of two first-order operators, a method analogous to Dirac’s approach to solving the harmonic oscillator \cite{b45be55d-49ab-3799-9fe1-69b428bbfb51}. Subsequently, it was recognised that SQM is related to this factorisation method, which had been categorised in 1951 by Infeld and Hull \cite{Infeldhull}. Notably, Gendenshtein (1983) introduced the concept of a \textit{Shape-Invariant Potential} (SIP) \cite{Gendenshtein:1983skv}: supersymmetric partner potentials sharing the same spatial dependence but differing in parameters. A key property is that the exact energy eigenvalues for these potentials can be obtained purely algebraically. SIPs have subsequently been applied to well known quantum mechanics problems, such as the harmonic oscillator and the Coulomb potential, including the hydrogen atom \cite{SISSAKIAN1990247,PhysRevA.48.1089,Markovich:2011vjm}. Supersymmetric Hamiltonians have also been studied in \cite{largeNexp,david123124124123}.

The Supersymmetric WKB (SWKB) approximation was proposed as a semiclassical method \cite{COMTET1985159} that incorporates the algebraic structure underlying supersymmetric quantum mechanics (SQM). Its principal utility is in treating shape-invariant potentials (SIPs) \cite{gangotundoublewell,SIL1994209}. For these systems, under standard assumptions, the leading-order SWKB quantisation condition produces the \textit{exact} energy eigenvalues. Consequently, the formalism constitutes a natural semiclassical realisation of the broader SQM framework, enabling direct determination of the energy spectrum \cite{Cooper_1995}.

While its primary strength lies at the spectral level, it can be complemented by supersymmetric constructions or standard WKB methods to obtain wavefunctions and related physical quantities such as tunnelling amplitudes. In practice, the SWKB approach typically begins by determining the energy spectrum via the corresponding quantisation condition, which, being exact for shape-invariant potentials \cite{DELANEY1990301,2009venn}, yields the classical turning points accordingly. These points can then be used as input to standard semiclassical constructions of the wavefunction, thereby providing a consistent description of the phase and barrier structure. The supersymmetric superpotential further refines the semiclassical action by incorporating exact ground-state information, often leading to improved estimates of tunnelling amplitudes compared to standard WKB, particularly in shape-invariant systems. Previous studies have compared supersymmetric and standard WKB approximations for several quantum-mechanical models \cite{SIL1994209}, finding, in several cases, improved estimates of tunnelling probability within the SWKB framework. These conclusions were possible in particular cases that enabled the determination of non-trivial  \textit{exact} expressions for the tunnelling probabilities. Related results on barrier penetration for shape-invariant potentials, obtained via operator methods or exact calculations of tunnelling probabilities, can be found in \cite{AKhare_1988,Aleixo_2000}. On this basis, the present work was developed.

Concretely, the standard method for treating tunnelling in quantum cosmology is the WKB approximation \cite{pmonizbook,monizQuantumCosmology2,kouniatalisWaveFunctionUniverse2025}.
Briefly, a minisuperspace approximation \cite{Kiefer2004_KIEQG,socorro,wvepacketsmini, pmonizbook, monizQuantumCosmology2,kouniatalisWaveFunctionUniverse2025,bohmian} is adopted \cite{PhysRevD.33.3560}. In particular, the Wheeler-DeWitt equation can be reduced to a one-dimensional second-order differential equation, permitting the application of standard quantum-mechanical
tunnelling methods.

A well-motivated scenario to investigate is the transition, in a Friedmann-Robertson-Walker (FRW) model with a cosmological constant, from a matter-dominated to a dark-energy-dominated state induced by a generalised Chaplygin gas potential barrier \cite{alternativequintessence,bentoGeneralizedChaplyginGas2002}. A detailed analysis and results are presented in \cite{Bouhmadi_L_pez_2005}.

A plausible objective for further study is therefore to perform a
careful comparison between the WKB and SWKB methods within a quantum
cosmological setting derived from \cite{Bouhmadi_L_pez_2005}. Although supersymmetry has long been applied in quantum cosmology \cite{socorro,pmonizbook,monizQuantumCosmology2}, to our knowledge, the application of the SWKB approximation to minisuperspace quantum cosmology remains underexplored \cite{Kiefer2004_KIEQG}.

The minisuperspace model is presented in Section \ref{minisuperspace_sec}. A brief introduction to SQM in the context of the minisuperspace model is provided in Section \ref{section_susy}. In Section \ref{approx_sec} analytical approximations (namely, power-series solution and the Picard method) are applied to solve the Riccati ordinary differential equation and obtain the superpotential. Both approximations indicate that the SQM potential produces qualitatively different late-time behaviour compared to the standard Chaplygin gas, where the characteristic negative slope for large scale factor values in the latter is replaced by a well, featuring a significant “wall” in the former. In quantum cosmology, the tunnelling probability is interpreted as the likelihood of the universe transitioning into a specific evolutionary phase. A backward numerical integration method is introduced as a benchmark, confirming that the true transmission probabilities are strictly bounded by the two semiclassical approximations. This demonstrates that standard WKB systematically underestimates the tunnelling probability, while SWKB overestimates it, which may be attributed to the lack of shape-invariance. Crucially, by confirming that the quantum transition through the potential barrier is much more probable than standard WKB estimates suggest, these results firmly support the physical viability of transient late-time acceleration models. A discussion and conclusion on the obtained results is presented in Section \ref{conclusion}. A functional form for the resulting decaying dark energy density is proposed as a next step for work. This would allow to appraise the relevance of a SQM-induced minisuperspace potential, against current observational data. Additionally, an appendix detailing the covariant derivation of the minisuperspace Lagrangian and clarifying the conceptual distinction between the Wheeler--DeWitt and standard Schrödinger equations is provided. A brief appendix on SQM applied to the harmonic oscillator is included. Another appendix illustrates the SWKB procedure for a typical SQM tunnelling within SIP: the Pöschl--Teller case.
These standard, exactly solvable examples are included strictly to provide a pedagogical baseline and technical intuition for the SWKB methodology. The rigorous consistency tests validating the application of SWKB to our specific, non-shape-invariant Chaplygin gas model are performed independently via backward numerical integration (\cref{sec:tun_prob_calc}) and higher-order analytical bounds (\ref{3rd_Picard}).

\section{Minisuperspace model} \label{minisuperspace_sec}
Consider the minisuperspace model employed in \cite{Bouhmadi_L_pez_2005} that describes a closed Friedmann-Robertson-Walker (FRW) universe with a generalised Chaplygin gas. The minisuperspace Lagrangian is given by \cite{Bouhmadi_L_pez_2005,Hawking_Ellis_1973}
\begin{equation}\label{eq:larangian1}
    L=- \frac{3\pi}{4G}\left(\frac{\dot{a}^2a}{N}-Na \right)-2\pi^2 a^3N \rho.
\end{equation}
 
Here, $a$ is the scale factor, $G$ is Newton's constant, $\rho$ is the energy density of the generalized Chaplygin gas defined ahead in equation \eqref{fullchap}, and $N$ is the lapse function arising from the $3{+}1$ (ADM) decomposition. A detailed derivation of this minisuperspace Lagrangian from the full covariant action is provided in \ref{app:action_wdw}. Variation of the action with respect to $N$ enforces the Hamiltonian constraint 
\begin{equation} \label{eq:hami_const} 
\frac{G}{3\pi}\pi^2_a+\frac{3\pi}{4G}\left(a^2 - \frac{8\pi G}{3}a^4 \rho\right)=0,
\end{equation}
where $\pi_a={\partial L}/{\partial\dot{a}}=-\tfrac{3\pi}{2G}\,a\dot a/N$ is the canonically conjugate momentum to $a$. By rescaling the energy density for the Chaplygin gas $\bar{\rho}(a) \equiv \frac{8\pi G}{3} \rho(a)$, we may write this equation as
\begin{equation} 
   \frac{G}{3\pi}\pi^2_a+\frac{3\pi}{4G}V(a)=0. 
\end{equation}
Here the potential $V(a)$ is defined by the expression
\begin{equation}\label{fullchap}
    V(a)=a^2-a^4\bar{\rho}(a) \,, \quad \quad \quad  \bar{\rho}(a)=\left( \bar A+\dfrac{\bar B}{a^{3(\alpha+1)}}\right)^{\frac{1}{1+\alpha}},
\end{equation}
where $0< \alpha \leq 1 $, $\bar A=A\left(\frac{8\pi G}{3}\right)^{1+\alpha}$ and $\bar B=B\left(\frac{8\pi G}{3}\right)^{1+\alpha}$, with $A, \,B$ being positive constants. For $\alpha=1$, we recover the usual Chaplygin gas case.
This model's quantum description is obtained by quantizing the conjugate momenta $\pi_a \to -i \frac{d}{da}$ to obtain the Wheeler-DeWitt equation
\begin{equation}
\left[-\dfrac{G}{3\pi}\dfrac{d^2}{da^2}+ \dfrac{3\pi}{4G}V(a)\right]\Psi(a)=0\label{wdw1},
\end{equation}
where the wavefunction $\Psi(a)$ represents the quantum state of the universe. Note that in equation \eqref{wdw1} factor ordering ambiguities were neglected, since our subsequent analysis relies on semiclassical approximations to calculate transmission probabilities. Additionally, the unit convention $2G=3\pi$ is adopted, simplifying the Wheeler-DeWitt equation
\begin{equation}
\left[-\dfrac{d^2}{da^2}+ V(a)\right]\Psi(a)=0.\label{wdw2}
\end{equation}
This equation formally resembles a Schrödinger equation, with a zero-energy eigenvalue. However, due to the timelessness nature of the Hamiltonian constraint, they are physically distinct. Further details on this distinction are provided in \ref{app:action_wdw}.
\begin{figure} 
    \centering
    \includegraphics[width=0.5\linewidth]{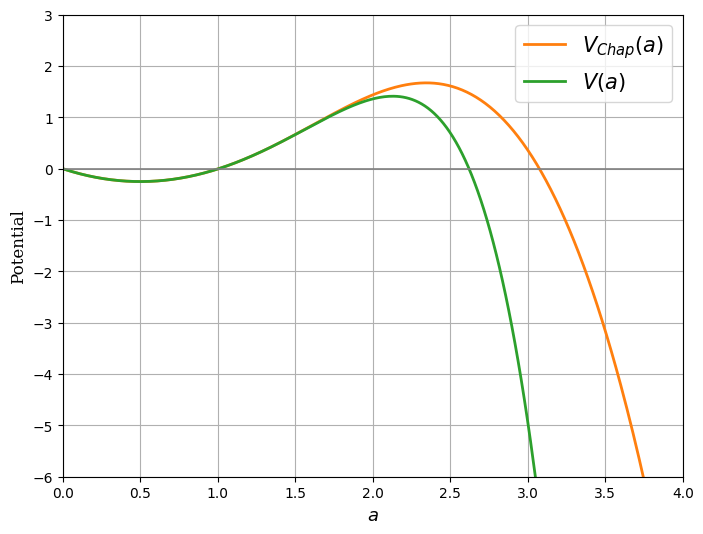}
    \caption{Comparison of the Generalised Chaplygin gas potential (orange), and its approximation (green), for $A=0.01$, $B=1$ and $\alpha=1$. The main features of the  Chaplygin gas potential, the initial well, the barrier and the sharp decay, are preserved in this approximation.}
    \label{chapapprox}
\end{figure}
Following \cite{Bouhmadi_L_pez_2005}, the energy density $\bar \rho(a)$ is expanded for small values of $a$, yielding (where we henceforth drop the overbars on A and B)
\begin{equation}
    \left( A+\dfrac{B}{a^{3(\alpha+1)}}\right)^{\frac{1}{1+\alpha}}\approx \dfrac{B^\frac{1}{1+\alpha}}{a^3} \left[ 1+ \frac{1}{1+\alpha}\frac{A}{B}a^{3(\alpha+1)} - \frac{\alpha}{2(1+\alpha^2)} \frac{A^2}{B^2}a^{6(\alpha+1)}\,+\,...\right]. \label{expansion_chap}
\end{equation}
This expansion is valid for
\begin{equation}
    a \ll \left((1+\alpha)\dfrac{B}{A}\right)^{\frac{1}{3(1+\alpha)}}.
\end{equation}
To leading order, the minisuperspace potential for the generalised Chaplygin gas approximates to (\cref{chapapprox})
\begin{equation}
    V(a)\approx a^2-aB^\frac{1}{1+\alpha}-\dfrac{1}{1+\alpha}\frac{A}{B^\frac{\alpha}{1+\alpha}}a^{3\alpha+4} \label{pot_desenvolv}.
\end{equation}


It should be stressed that the expansion leading to Eq.\eqref{pot_desenvolv} is used \emph{only} to describe the quantum-mechanical tunnelling of the wavefunction of the universe from the effective dust-dominated phase, through the potential barrier, into the region of unbounded classical expansion -- a restriction to small $a$ that is standard practice in this class of models \cite{Kiefer2004_KIEQG,alternativequintessence,Bouhmadi_L_pez_2005}. Any discussion regarding the subsequent dark-energy-dominated accelerating phase of the \emph{standard} model refers to the \emph{classical} evolution of the full generalized Chaplygin gas, governed by Eq.\eqref{eq:hami_const}--\eqref{fullchap}, and not to an extrapolation of the small-$a$ potential \eqref{pot_desenvolv} outside its domain of validity. In the unmodified (non-SQM) model this classical late-time evolution can, for certain subclasses of the generalized Chaplygin gas, approach a curvature singularity at $a\to\infty$ of the sudden/future type \cite{barrowSuddenFutureSingularities2004a}, as already discussed in \cite{Kiefer2004_KIEQG,alternativequintessence,Bouhmadi_L_pez_2005}; this generic feature is independent of, and unaffected by, the small-$a$ quantum treatment used here. In contrast, as demonstrated below, applying the structural criteria of SQM fundamentally alters this late-time picture, replacing the unbounded expansion with a transient accelerating phase that is structurally prevented from persisting indefinitely (see \cref{picardsection} for a detailed discussion of this feature).

As may be observed in \cref{chapapprox}, the potential \eqref{pot_desenvolv} exhibits the same behaviour as the generalised Chaplygin gas: an initial bound region corresponding to the matter-dominated era, followed by a barrier, and subsequently a descending slope associated with the dark-energy-dominated era. For brevity, we henceforth refer to \eqref{pot_desenvolv} as the Chaplygin potential $V(a)$. The parameter $A$ still serves as an effective cosmological constant, controlling the large $a$ behaviour, consistent with the full Chaplygin gas potential \eqref{fullchap}.

\section{Supersymmetric Quantum Mechanics}\label{section_susy}
Supersymmetric Quantum Mechanics (SQM) is a robust framework that, for systems that permit the factorisation of the Hamiltonian into two first-order differential operators, facilitates the calculation of energy spectra for bound systems. It is particularly effective for analysing the system's ground state and computing tunnelling probabilities. SQM has been applied across various physics domains, including solid-state and nuclear physics \cite{Cooper_book}. Applications include canonical problems such as the harmonic oscillator (with a ladder operator method analogous to that presented here, and a SUSY harmonic oscillator example in  \ref{app:susyho}), the Hydrogen atom, among others \cite{PhysRevA.48.1089,Cooper_book}. Let us remark before we proceed that we are strictly employing SQM as formulated in references such as \cite{Cooper_book,Cooper_1995,gangopadhyaya2024} and that any extensions that may include the presence of Grassmannian variables (see e.g., \cite{Lidsey1995Duality,LidseyMaharana1998,Lidsey1995PRD,LidseyMoniz2000,Junker1996,pmonizbook,monizQuantumCosmology2}) will not be considered in this work.
The factorisation operators are defined as follows.
\begin{align}
    A&\equiv\dfrac{d}{da}+W(a) \notag \\
    A^\dagger&\equiv-\dfrac{d}{da}+W(a).
\end{align}
Here, the superpotential $W(a)$ is introduced as a continuous and differentiable function of $a$. Two isospectral partner Hamiltonians are defined by composing these operators:
\begin{align} \label{factorization}
   H_+ &\equiv  AA^\dagger=-\dfrac{d^2}{da^2}+V_+(a), \notag \\
   H_- &\equiv A^\dagger A=-\dfrac{d^2}{da^2}+V_-(a). \notag \\
\end{align}
The potentials $V_\pm$ are defined as
\begin{equation}
    V_\pm(a)\equiv W^2(a) \,\pm \dfrac{dW}{da}. \label{supot_eq}
\end{equation}
These Hamiltonians are positive definite, implying that all energy eigenvalues are positive except for a unique zero-energy ground state associated with one Hamiltonian \cite{land2024supersymmetric}. When this condition holds (noting that only one partner Hamiltonian possesses a zero-energy ground state), the system exhibits \textit{unbroken} supersymmetry. Conversely, if neither Hamiltonian has a zero-energy ground state, the supersymmetry is considered \textit{broken} \cite{Gangopadhyaya_book}.

The ground state wavefunction of ${H}_\pm$ is given by
\begin{equation}
    \psi_0^\pm (x)\propto e^{\pm\int^a W(s)\,ds }.
    \label{eq-gstate-1}
\end{equation}
The normalizability of the ground state wavefunction depends on the asymptotic behaviour of $W(a)$. Restricting our domain to $a \geq 0$, we require that $\int^\infty W(s)\, ds \to \mp \infty$ to ensure the decay of $\psi_0^\pm$, respectively, at large $a$. Consequently, at most only one partner Hamiltonian may possess a normalizable zero-energy ground state \cite{Gangopadhyaya_book}.

In this context, as $a$ is restricted to positive values, the normalisation condition is imposed over the positive real axis:
\begin{equation}
    \int _{0}^{+\infty} |\psi_0(a)|^2\,da = 1. \end{equation}
To ensure normalizability of $\psi_0^\pm$, the behaviour of the superpotential as $a \to +\infty$ is critical, assuming no divergences occur at zero or other points.

The SQM framework is originally formulated for the Schrödinger equation, whereas \cref{minisuperspace_sec} discusses the fundamentally different Wheeler-DeWitt equation. However, as was already discussed in the previous section and in \ref{app:action_wdw}, the equation under consideration has a similar structure when the Schrödinger equation is imposed with $ E=0$. Therefore, applying the SQM formalism (as in \cite{Cooper_book,Gangopadhyaya_book,WITTEN1981513,Cooper_1995,COOPER1983262}) to the minisuperspace model is well-motivated.
This approach relies on the zero-energy condition ($E=0$) enforced by the Wheeler-DeWitt equation, which allows us to formally define a superpotential and directly evaluate the zero-energy states, regardless of whether these solutions are normalizable (unbroken supersymmetry) or non-normalisable (broken supersymmetry).

\section{Approximated Superpotentials} \label{approx_sec}

The primary challenge in applying the supersymmetric framework is to determine an analytical superpotential $W(a)$ for a general potential. This requires solving the non-linear differential equation (\ref{supot_eq}). Either the plus or minus sign can be used, since if $W(a)$ is a solution to \eqref{supot_eq} with the minus sign, then $-W(a)$ is a solution to the equation with the plus sign. The positive sign is adopted for the subsequent analysis. If we are to obtain the superpotential, we need to solve 
\begin{equation} \label{eq:diff_eq_V} 
    W^2(a) + \frac{dW}{da} = V(a) = a^2-aB^\frac{1}{1+\alpha}-\dfrac{1}{1+\alpha}\frac{A}{B^\frac{\alpha}{1+\alpha}}a^{3\alpha+4}.
\end{equation}
Since there is no \emph{exact} analytical solution for this equation, we say the potential $V(a)$ \eqref{pot_desenvolv} cannot be \emph{exactly} factorized, since we cannot directly write the Hamiltonian as a product of two linear operators as in equation \eqref{factorization}. Therefore, approximate methods are necessary to obtain $W(a)$.

In standard Supersymmetric Quantum Mechanics (SQM) for bound systems, the superpotential is constructed to shift the energy scale. Specifically, the partner potential is defined as $V_+(a) = V(a) - E_0$, yielding a shifted Hamiltonian $H_+(a) = H(a) - E_0$. This guarantees that the ground state of $H_+(a)$ possesses exactly zero energy, as demonstrated in \ref{app:susyho}.

However, our minisuperspace model is not a completely bound system, and the Wheeler-DeWitt equation, $\mathcal{H}\Psi(a)=0$, intrinsically imposes a zero-energy eigenvalue. Consequently, no energy shift is required. The superpotential must instead be constructed so that the resulting SQM Hamiltonian directly approximates the original Hamiltonian constraint itself, such that $\mathcal{H}(a) \approx H_+(a)$.

\subsection{Power-Series method}

We proceed to obtain an approximate solution in the simplest case ($\alpha=1$), where the potential $V(a)$ becomes a polynomial
\begin{equation}
    V(a)=a^2-\sqrt B a - \frac{Aa^7}{2\sqrt B}. 
\end{equation}
Since, by imposing this condition, we are working with integer powers, we may adopt a polynomial ansatz to solve equation \eqref{eq:diff_eq_V}. We use a power-series superpotential of degree $N$:
\begin{equation}
    W_\text{ps}^{(N)}(a)= \sum_{n=0}^N w_na^n.    
\end{equation}
This superpotential ansatz is then substituted into equation \eqref{eq:diff_eq_V}, and terms of the same order in $a$ are matched. We impose the condition $W(0)=0$, which sets $w_0=0$. The resulting equation is
\begin{align}
    V(a)=a^2-a\sqrt B - \frac{A}{2\sqrt B}a^7 = & \left(\sum_{j=1}^N w_ja^j\right)\left(\sum_{k=1}^N w_ka^k\right) + \frac{d}{da}\left(\sum_{k=1}^N w_ka^k\right) \\ \notag
     =& \sum_{j=1}^N \sum_{k=1}^N w_j w_k a^{j+k} + \sum_{k=1}^N k w_k a^{k-1}. \label{matchcoef_eq}
\end{align}
\begin{figure}
    \centering
    \includegraphics[width=0.5\linewidth]{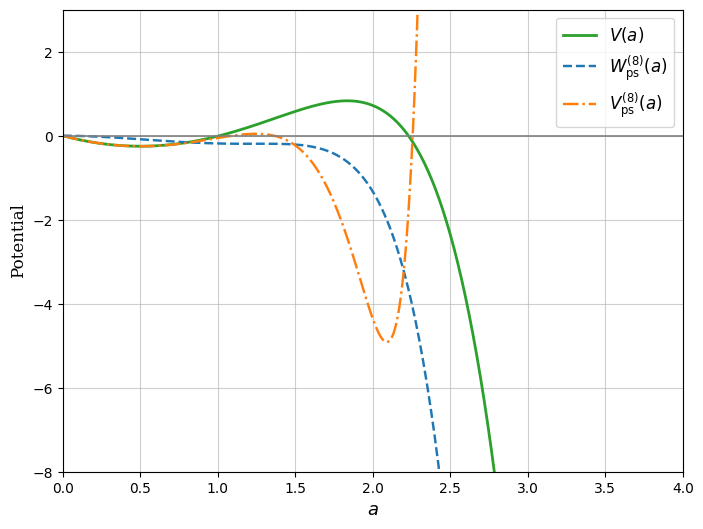}
    \caption{Comparison of the 8th-order power-series superpotential $W_\text{ps}^{(8)}(a)$ (blue), the approximated potential $V_\text{ps}^{(8)}(a)$ (orange), and the exact Chaplygin potential $V(a)$ (green), evaluated at $B=1$, $\alpha=1$, and $A=0.02$. The SQM approach accurately reproduces the exact potential for small $a$. However, for large $a$, rather than decaying, $V_\text{ps}^{(8)}(a)$ turns upward to form a confining wall, a generic consequence of polynomial superpotentials.
}
    \label{plotpower1}
\end{figure}
Matching coefficients in $a$ up to $a^7$ yields
\begin{align} \notag
    \text{Order $a^0$:}& \quad w_1=0 \\[3pt] \notag
    \text{Order $a^1$:}& \quad 2w_2= -\sqrt B \\[3pt] \notag
    \text{Order $a^2$:}& \quad w_1^2+3w_3=1 \\[3pt]
    \text{Order $a^3$:}& \quad 2w_1w_2+4w_4=0 \\[3pt] \notag
    \text{Order $a^4$:}& \quad 2w_1w_3+w_2^2+5w_5=0\\[3pt] \notag
    \text{Order $a^5$:}& \quad 2w_1w_4+2w_2w_3+6w_6=0\\[3pt]\notag
    \text{Order $a^6$:}& \quad 2w_1w_5+2w_2w_4+w_3^2+7w_7=0\\[0pt]\notag
    \text{Order $a^7$:}& \quad 2w_1w_6 + 2w_2w_5 + 2w_3w_4 + 8w_8 = -\frac{A}{2\sqrt{B}}\\\notag
\end{align}
Although, at first glance, this appears to be a system of coupled non-linear algebraic equations, we notice that at the $n$-th order, the coefficient $w_{n+1}$ appears isolated and depends only on lower-order coefficients. By writing $V(a)=\sum_n v_n a^n$ -- where $v_1=-\sqrt{B}$, $v_2=1$, $v_7=-A/(2\sqrt{B})$ and $v_m=0$ otherwise -- we are able to write
\begin{equation}
    v_n=\sum_{j=1}^N\sum_{k=1}^N w_j w_k\delta_{j+k,n}+ \sum_{k=1}^N kw_k\delta_{k-1,n},
\end{equation}
and by summing over the Kronecker deltas, we obtain a recursive relation for $w_{n+1}$
\begin{equation} \label{eq:recursive_pow}
    w_{n+1}=\frac{1}{n+1}\left(v_n-\sum_{k=1}^{n-1}w_kw_{n-k}\right). 
\end{equation}
This recursive formula allows us to determine the next coefficient by summing over products of previously obtained ones. It is important to point out that in this equation, the summation runs from $k=1$ to $k=n-1$. This is because we set the condition $w_0=0$. If a different initial condition was imposed, we would sum from $k=0$ to $k=n$ (since $w_0=0$, here we could technically still sum from $k=0$ to $k=n$ because these additional terms would vanish). From this recursion relation, we obtain the 8th-order superpotential:
\begin{figure}
    \centering
    \includegraphics[width=0.5\linewidth]{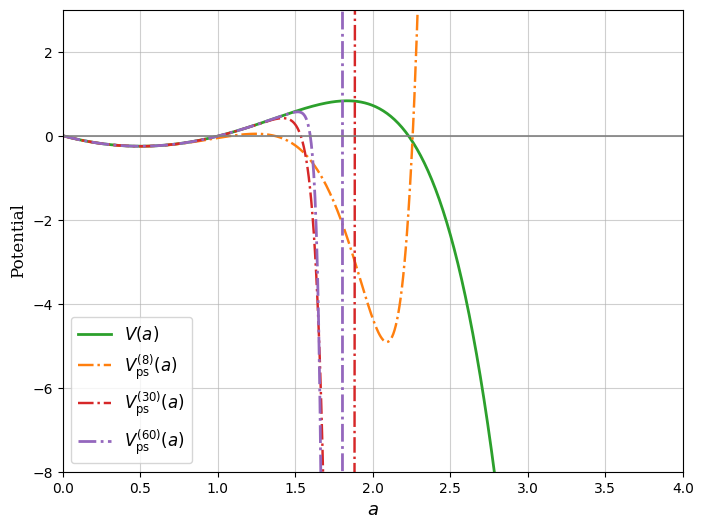}
    \caption{Comparison of the power-series-generated potentials at higher orders of approximation. Shown are the 8th-order $V_\text{ps}^{(8)}(a)$ (orange), 30th-order $V_\text{ps}^{(30)}(a)$ (red), and 60th-order $V_\text{ps}^{(60)}(a)$ (purple) approximations alongside the exact Chaplygin potential $V(a)$ (green), evaluated at $B=1$, $\alpha=1$, and $A=0.02$. Despite the significantly increased order of the expansion, the approximations fail to capture the decaying asymptotic tail. Instead, they consistently turn upward at large $a$, confirming that the confining wall is an inescapable feature of the polynomial ansatz.
}
    \label{fig:match_power}
\end{figure}
\begin{equation}\label{eq:powerseriessuperpot}
    W_\text{ps}^{(8)}(a)=-\frac{\sqrt{B}}{2}a^2 + \dfrac{1}{3}a^3-\frac{B}{20}a^5+\frac{\sqrt{B}}{18}a^6-\frac{1}{63}a^7-\frac{(10A+B^2)}{160\sqrt B}a^8.
\end{equation}
Substitution into \eqref{supot_eq} yields the approximated potential
\begin{equation}
V_\text{ps}^{(8)}(a) = a^2 - \sqrt{B}a - \frac{A a^7}{2 \sqrt{B}} + \mathcal{O}(a^8).
\end{equation}
Due to the derivative term in the Riccati equation, an $N$-th order superpotential matches the exact potential up to order $N-1$. It is important to note that this approximation is valid only for small $a$, as is clearly illustrated in \cref{plotpower1}. While $V_\text{ps}^{(8)}(a)$ exactly reproduces the Chaplygin potential up to $a^7$, the residual $\mathcal{O}(a^8)$ terms, which include even powers up to $a^{16}$ generated by the $W^2(a)$ term, dominate at large $a$, forming the confining wall. The plus sign in equation \eqref{supot_eq} was selected so that the superpotential behaves as $W(a) \to -\infty$ when $a \to +\infty$, which is visually more intuitive. If the minus sign in \eqref{supot_eq} were chosen instead, the resulting superpotential would simply be inverted along the $E=0$ axis. As shown in \cref{plotpower1}, the superpotential $W_\text{ps}^{(8)}(a)$ approaches $-\infty$ for large $a$, implying the existence of a normalizable ground state $\psi_0^+ \propto \exp \int^a W(s) \, ds$.

In an attempt to obtain a better approximation, we employ a higher-order superpotential ansatz and the recursive relation \eqref{eq:recursive_pow}. Using a 60th-order superpotential, we still find that the barrier is well approximated for small values of $a$. As seen in \cref{fig:match_power}, this method converges slowly. It is important to stress that, while calculating the superpotential up to 60th order, we strictly used the exact 7th-order polynomial form of $V(a)$ for $\alpha=1$. This choice was deliberate to isolate the convergence behavior of the superpotential ansatz itself, without introducing higher-order expansion terms from the original Chaplygin potential. An identical analysis could be performed using a higher-order expansion in \eqref{expansion_chap}, yielding a similar result in which the approximation scheme is still only valid for small $a$. While we document this methodology here as it may prove effective for other classes of potentials in future studies, its slow convergence for our specific case prompts us to search for an alternative approximation method. As discussed in the next subsection, this alternative yields far better results even at first order.

\subsection{Picard's method}\label{picardsection}

\begin{figure}
    \centering
    \includegraphics[width=0.5\linewidth]{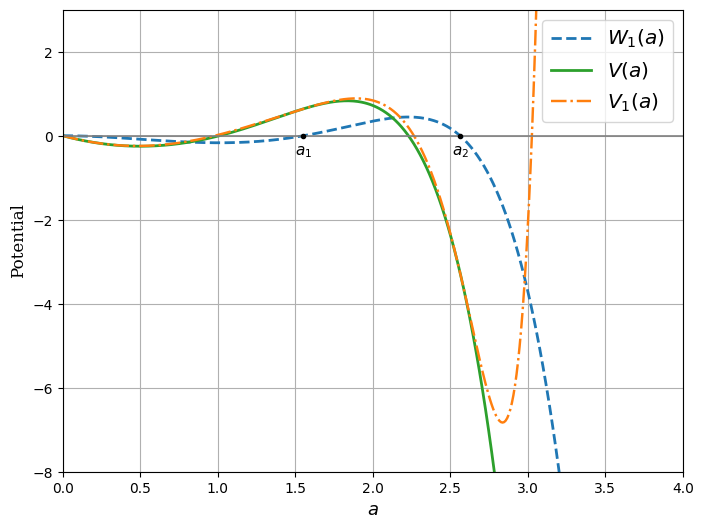}
    \caption{Plot of the first iteration Picard superpotential $W_1(a)$ (blue), the approximated potential $V_1(a)$ (orange) and the Chaplygin potential $V(a)$ (green), with $B=1, \,\,\,\, \alpha=1, \,\,\,\, A=0.02$. The classical turning points for the superpotential ($W^2(a_{1,2})=0$) are also plotted. Compared with the previous approximation, this one is better behaved near the barrier for the same parameter values. However, for large $a$, the approximate potential differs qualitatively from the original one:  whereas $V(a)$ decays, $V_1(a)$ has an inflexion point and starts to rise sharply, forming a confining wall. This is a generic feature of polynomial superpotentials, as discussed in Section \ref{swkb_section}.}
    \label{fig:3plots1}
\end{figure}

While the power-series approach provides a valid approximation for sufficiently small $a$, its nature as a local expansion restricts its ability to reliably reproduce the entire potential barrier across a broader parameter range. Furthermore, the coefficient-matching procedure explicitly requires restricting the gas parameter $\alpha=1$. To overcome these limitations, we seek a more versatile analytic approximation via Picard's method. First, we express equation \eqref{supot_eq} as
\begin{equation}
   \frac{dW}{da}=V(a)-W^2(a)=a^2-aB^\frac{1}{1+\alpha}-\dfrac{1}{1+\alpha}\frac{A}{B^\frac{\alpha}{1+\alpha}}a^{3\alpha+4} - W^2(a),
\end{equation}
Thus, the first approximation $W_1(a)$ is obtained by integrating on both sides:
\begin{equation}
W_1(a)= W_0+\int_{0}^a V(s)- W_0^2 \,ds,
\end{equation}
Here, $s$ is just an integration variable, and the initial condition $W_0\equiv W(a_0=0) = 0$ is imposed. Substituting the potential and performing the integration yields
\begin{align}
W_1(a)&=\int_{0}^a V(s)\,ds \notag \\ 
    &=\int_{0}^as^2-sB^\frac{1}{1+\alpha}-\dfrac{1}{1+\alpha}\frac{A}{B^\frac{\alpha}{1+\alpha}}s^{3\alpha+4} \,ds \notag \\ 
    &= \dfrac{a^3}{3}-\dfrac{a^2}{2} B^\frac{1}{1+\alpha} - \dfrac{1}{1+\alpha}\frac{A}{B^\frac{\alpha}{1+\alpha}}\frac{a^{3\alpha+5}}{3\alpha+5}, \label{eq_super_w1}
\end{align}
This superpotential $W_1(a)$ is plotted in \cref{fig:3plots1}. The subsequent iteration is easily obtained by replacing $W_0(a)$ with $W_1(a)$ and $W_1(a)$ with $W_2(a)$. This can be generalized in the recursive relation
\begin{equation}
    W_{n+1}(a)=\int_0^a V(s)-W_n^2(s)\,ds. 
\end{equation}
We may simplify this expression by recognizing $\int_0^aV(s)\,ds=W_1(a)$
\begin{equation}\label{eq:pic_norder}
    W_{n+1}(a)=W_1(a)-\int_0^aW_n^2(s)\,ds. 
\end{equation}
The primary limitation of this iterative approach is that the degree of the polynomial grows rapidly with each step. While the integration itself relies entirely on the elementary power rule, making higher-order iterations mechanically straightforward, the resulting algebraic expressions quickly become analytically intractable. For this reason, we restrict our subsequent analysis to the first-order approximation and relegate the explicit calculation of the second and third-order superpotential to the \ref{3rd_Picard}.
The validity of the first iteration \eqref{eq_super_w1} requires the derivative $W_1'(a)$ to dominate over $W_1^2(a)$, ensuring that $W_1(a)$ provides a reliable approximation to $W(a)$. This condition can be locally expressed as
\begin{equation}
    W_1^2(a)\ll |W_1'(a)|, 
\end{equation}
which, in our case, holds when $a$ is small. For large $a$, the quadratic term will dominate, causing the Picard approximation to lose accuracy. This establishes a natural boundary for the validity of the analytic results and motivates extending the analysis by numerically integrating equation \eqref{supot_eq}, for example, via a Runge-Kutta scheme. Such extensions are reserved for future work.

This superpotential yields a formally normalizable ground state for $H_+$:
\begin{equation}
    \psi_0^+(a) \propto \exp\left(\int^a W(s)\,ds\right),
\end{equation}
since $W_1(a)$ approaches $-\infty$ as $a  \to +\infty$.  While standard SQM strictly restricts the physical state to this normalizable ground state, the Wheeler--DeWitt equation does not necessarily require standard $L^2(\mathbb{R}^+)$ normalizability \cite{Kiefer2004_KIEQG, wiltshireIntroductionQuantumCosmology2003}. Consequently, in the classically forbidden region, its general solution will be given by a linear combination of the formal zero-energy solutions of both partner Hamiltonians $H_\pm$, 
\begin{equation}
    \Psi(a) \approx C_+ \psi_0^+(a) + C_- \psi_0^-(a),
\end{equation}
although, in our case, $\psi_0^-(a)$ is non-normalizable. In quantum cosmology, both of these mathematical solutions carry profound physical meaning.

Specifically, in standard WKB, the sub-barrier wavefunction is governed by the integral of the semiclassical momentum, $K_{{WKB}}(a) = \int^a \sqrt{V(s)} \, ds$. The SWKB framework dramatically simplifies this structure by replacing the momentum with the analytic superpotential we just derived. Thus, the local SWKB exponent is defined purely by integrating the superpotential: $K_{{SWKB}}(a) = \int^a |W(s)| \, ds$. As we will detail in \cref{swkb_section}, evaluating the definite integral of this function between the appropriate turning points yields the total tunnelling exponent $K_{{SWKB}}$, from which the tunnelling probabilities are computed. Since $W_1(a)$ is strictly positive within the integration domain bounded by the turning points (plotted in \cref{fig:3plots1}), the normalizable SQM ground state $\psi_0^+(a) \propto \exp\left(+\int^a W(s)\, ds\right)$ corresponds to the growing SWKB branch $\exp(+K_{SWKB}(a))$, effectively aligning with the Hartle--Hawking (no-boundary) wavefunction \cite{hawkinghartlewfofuniverse}. Conversely, the non-normalizable formal solution $\psi_0^-(a) \propto \exp\left(-\int^a W(s)\, ds\right)$ matches the decaying SWKB branch $\exp(-K_{SWKB}(a))$. This would correspond to the Vilenkin (tunnelling) wavefunction \cite{vilenkinQuantumCreationUniverses1984}.

The SQM-generated potential is given by, setting $\alpha=1$ for simplicity
\begin{align} \label{potsqm1}
    V_1(a) \equiv V_+(a)= & \,W_1^2+\frac{dW_1}{da}= \\ \notag
    = & \,\left(a^2-a\sqrt{B}-\frac{a^7A}{2\sqrt{B}}\right) - \frac{a^5\sqrt{B}}{3}+\frac{a^4B}{4}    +  \\ \notag
    &+ \frac{a^6}{9}+\frac{a^{10}A}{16}-\frac{a^{11}A}{24\sqrt{B}}+\frac{a^{16}A^2}{256B},
\end{align}
The first three terms correspond to the potential we intend to approximate. In the original potential $V(a)$, the barrier height is directly determined by $A$: smaller values of $A$ correspond to larger barriers. However, in $V_1(a)$, the presence of lower-order terms independent of $A$ implies that for smaller $A$, these powers dominate over the seventh-order one in the original potential. Consequently, the approximation deteriorates for small $A$.

The approximated potential diverges from $V(a)$ for large $a$, a behaviour inherent to potentials derived from polynomial superpotentials, as dictated by equation \eqref{supot_eq}. Specifically, squaring a polynomial yields the highest-order \emph{positive} term that dominates the behaviour of the SQM-generated potential for sufficiently large $a$, when compared to its derivative. Therefore, $W^2(a) \gg W'(a)$ for sufficiently large $a$. This structural constraint resolves the apparent contradiction of inferring late-time dynamics from a superpotential derived via a small-$a$ expansion. Rather than improperly extrapolating the standard model's dust-dominated regime, the SQM framework generates a new, globally defined potential that inherently departs from unbounded expansion.

As mentioned in this section, any finite-degree polynomial superpotential yields $V(a) \sim W^2(a) \to +\infty$ as $a\to+\infty$. We can see that this is the case for the vast majority of SQM potentials, since $W^2(a)$ will dominate over $W'(a)$ for large $a$ in most cases. One can also obtain a potential that asymptotically goes to zero (or to a positive constant) for large $a$, for example, with a rational superpotential of the form $W_{rat}=(p_1a^3+p_2a^2)/(1+q_1a^6)$.

This behaviour is fundamentally rooted in the standard SQM formalism. As we have seen before, the SQM Hamiltonians factorize as $H_- = A^\dagger A$ and $H_+ = A A^\dagger$, as they are positive definite operators. This guarantees that their energy spectra are strictly bounded from below by zero ($E \ge 0$). Consequently, the corresponding SQM potential cannot plunge to $-\infty$ for large $a$, as this would allow for localized wavepackets at large scale factors with negative energy expectation values, strictly violating the positive definiteness of the Hamiltonian. In cosmological terms, this mathematical constraint dictates that the SQM framework structurally forbids the unbounded, accelerating dark energy tail of the standard Chaplygin gas, forcing the potential to either turn upward or plateau. While the specific feature of an infinite confining wall diverging to $+\infty$ is a mathematical consequence of employing a polynomial superpotential, the underlying physical constraint -- that the potential is bounded from below and cannot decay to $-\infty$ -- is an intrinsic, inescapable requirement of imposing supersymmetry. The cosmological implications of this constraint will be discussed in \cref{conclusion}.

Furthermore, the comparison of the SWKB and WKB quantization conditions for non-shape-invariant potentials has long been an object of study \cite{artkhare,Khare:2004kn}. The literature suggests that when analytical techniques are employed to construct the superpotential, the SWKB approximation outperforms the usual WKB in obtaining the energy spectrum \cite{miao, chakrabarti}. Although we employ an analytic superpotential in this paper, further research could focus on numerically solving the differential equation \eqref{supot_eq}. The application of alternative approximation schemes (Runge-Kutta, Galerkin) is left for future work.

\section{The Supersymmetric WKB approximation}\label{swkb_section}

The WKB (or WKBJ) approximation is a semiclassical approach to obtaining approximate wavefunctions, tunnelling probabilities, and eigenvalues of Hamiltonians with slowly varying potentials, compared to the system’s wavelength \cite{froman,Berera_Del_Debbio_2021}.
The Supersymmetric WKB (SWKB) approximation is a supersymmetry-inspired counterpart 
\cite{COMTET1985159,artkhare} that integrates the concepts of supersymmetric quantum mechanics (SQM) with the lowest-order WKB method.
It provides an exact quantisation condition for a class of shape-invariant potentials
 (SIPs), whereas standard WKB methods yield semiclassical wavefunctions and tunnelling
amplitudes.

The procedure is as follows:
\begin{enumerate}
\item Use the supersymmetric structure to define the superpotential $W(a)$ and the partner potentials, as given by equation (\ref{supot_eq}).
\item Determine the exact energy spectrum using the SWKB quantisation condition (see equation (\ref{quanticond_swkb}) below).
The turning points are defined by the condition $E = W^2(a)$.
\item Construct semiclassical wavefunctions using standard WKB expressions evaluated at the exact SWKB energies.
\item Compute tunnelling amplitudes using WKB barrier penetration integrals, optionally refined by incorporating the superpotential structure.
\end{enumerate}
This approach ensures exact spectral input while maintaining the flexibility of semiclassical wavefunction analysis.
Given a superpotential $W(a)$, the ground-state wavefunction of $V_-(a)$ is precisely as expressed in equation (\ref{eq-gstate-1}).
This expression provides an exact foundation for the semiclassical construction and guarantees correct asymptotic behaviour, which is essential for tunnelling problems.

In the case of unbroken supersymmetry for bound systems, the SWKB approximation yields accurate energy eigenvalues for large quantum numbers $n$ and is exact for the ground state ($n=0$), since $E_0=0$ is known \cite{Cooper_book,miao}. For shape-invariant potentials (SIP), a relation between the two partner potentials guarantees the exactness of this method \cite{Cooper_book,Hruska_1997, jalalzadehShapeInvariantPotentials2022}.
\begin{equation}
V_+(a,\lambda) = V_-(a, f(\lambda))+R(\lambda), \label{sip_cond}
\end{equation}
Here, $\lambda$ is a parameter, and $f(\lambda)$ and $R(\lambda)$ are arbitrary functions of this parameter. Notably, when the partner potentials satisfy this relation, it has been demonstrated in \cite{SIL1994209} that the resulting tunnelling probabilities are improved
(i.e., they are significantly closer to exact solutions when available) compared to those obtained using the standard WKB approximation; furthermore, the quantisation condition is exact \cite{Cooper_book,jalalzadehShapeInvariantPotentials2022}.
For unbroken supersymmetry, the SWKB quantisation condition is given by \cite{COMTET1985159,Sinha_2000}:
\begin{equation}
\int_{a_1}^{a_2} \sqrt{E_n-W^2(a)}\,da=\pi n, \label{quanticond_swkb}
\end{equation}
Here, $a_1$ and $a_2$ denote the classical turning points defined by $W^2(a_1) = W^2(a_2) = E_n$, and $n$ is an integer.

A similar expression to equation \eqref{quanticond_swkb} is employed to compute tunnelling probabilities \cite{Cooper_book,SIL1994209}, with a minus sign inside the square root since within the barrier $W^2(a) > E$. The SWKB tunnelling exponent is defined as
\begin{equation} \label{swkb_k} 
K_{SWKB}= \int_{a_1}^{a_2} \sqrt{W^2(a)-E}\,da \, \mp \int_{W(a_1)}^{W(a_2)}\frac{dW}{\sqrt{W^2-E}},
\end{equation}
where again $a_{1,2}$ represent the turning points $W^2(a_{1,2})=E$ (as plotted in \cref{fig:3plots1}). This can be compared with the standard WKB formula
\begin{equation}
K_{WKB}=\int_{b_1}^{b_2} \sqrt{V(a)-E}\,da,  \label{wkb_k}
\end{equation}
where similarly, $b_{1,2}$ are defined as $V(b_{1,2})=E$. These expressions are similar, however, the SWKB exponent involves the square of the superpotential instead of the potential itself. Tunnelling probabilities can be calculated using
\begin{equation}\label{tunprob}
T(E)=\frac{1}{1+e^{2K}},  
\end{equation}
where $K=K_{(S)WKB}$, depending on the method used. This expression is found to yield improved results near the top of the barrier compared to the conventional exponential form $e^{-2K}$ \cite{froman,SIL1994209}. For deep-tunnelling regimes, when $K \gg 1$, we find that $(1+\exp(2K))^{-1} \sim \exp(-2K)$.

The SQM potentials $V_1(a)$ and $V_\text{ps}(a)$ are not shape invariant, indicating they do not satisfy relation \eqref{sip_cond}. Consequently, the SWKB approximation is not guaranteed to be exact; nevertheless, it remains a well-founded semiclassical approximation under unbroken supersymmetry. This observation motivates the search for shape-invariant potentials in minisuperspace quantum cosmology \cite{jalalzadehShapeInvariantPotentials2022}. Since the Wheeler-deWitt equation is given by $\mathcal{H}\Psi(a) = 0$, we effectively have $E=0$, and the turning points must satisfy $W(a_1) = W(a_2) = 0$, which means the second integral in equation \eqref{swkb_k} vanishes, yielding the following expression for $K_{SWKB}$:
\begin{equation}
K_{SWKB}= \int_{a_1}^{a_2} |W(a)|\,da, \label{tunswkb_def}
\end{equation}
The tunnelling probability is then obtained by substituting this expression into equation \eqref{tunprob}. Because the superpotential is positive within the barrier, $|W(a)| = W(a)$. The WKB exponent for the potential $V(a)$ is given by
\begin{equation}
K_{WKB}= \int_{b_1}^{b_2} \sqrt{V(a)}\,da,
\end{equation}
while the WKB exponent for $V_1(a)$ is, naturally,
\begin{equation}
K_{WKB_1}= \int_{c_1}^{c_2} \sqrt{V_1(a)}\,da.
\end{equation}

\subsection{Tunnelling probabilities calculation}\label{sec:tun_prob_calc}

We are now equipped to calculate tunnelling probabilities using the SWKB, by means of integrating the previously obtained superpotentials from the first to the second turning point. Before doing so, it is important to notice that the superpotential $W_{p}(a)$ obtained by the power-series approach will not be a good approximation, as can be observed in \cref{plotpower1}. For this reason, we decided to work only with $W_1(a)$ and $V_1(a)$, the superpotential obtained from Picard's method \eqref{eq_super_w1} and the respective SQM potential \eqref{potsqm1}, since it provides a more reliable approximation to the potential $V(a)$ \eqref{pot_desenvolv}.

We compare the tunnelling probabilities calculated from the SWKB method as in equation \eqref{swkb_k}, using the approximated Picard superpotential $W_1(a)$ ($T_{SWKB}$), with the ones obtained from the WKB method as in equation \eqref{wkb_k} for the approximated potential $V_1(a)$ ($T_{{WKB}_1}$). Because this potential $V_1(a)$ is obtained from $W_1(a)$ by equation \eqref{potsqm1}, this is a direct comparison between the WKB and SWKB methods. We then calculate the WKB tunnelling probabilities using the Chaplygin potential $V(a)$ ($T_{WKB}$) and evaluate them against the probabilities obtained from the SQM potential $V_1(a)$. Since calculating tunnelling probabilities in the WKB scheme involves an integral over the (classical) barrier, similar probabilities correspond to a comparable barrier shape for $V(a)$ and $V_1(a)$. Because the turning points must be found numerically; Newton's method was used for this task, but the SWKB integral is evaluated analytically for our superpotential $W_1(a)$
\begin{align} 
    K_{SWKB}&= \int_{a_1}^{a_2} W_1(a)\,da \notag \\
    &=\left[\dfrac{a^4}{12}-\dfrac{a^3}{6} B^\frac{1}{1+\alpha} - \dfrac{1}{1+\alpha}\frac{A}{B^\frac{\alpha}{1+\alpha}}\frac{a^{3\alpha+6}}{(3\alpha+5)(3\alpha+6)}\right]_{a_1}^{a_2}. \label{eq:tunswkb_calc}
\end{align}
We calculate the tunnelling probabilities for the potentials $V(a)$ and $V_1(a)$ using the WKB method, evaluating the integrals \eqref{wkb_k} using Gaussian quadrature. 

\begin{figure}\centering 
\subfloat[Tunnelling probability $T(A,B,\alpha)$ versus $A$, with $B=0.8$ and $\alpha=1$;]{\label{a}\includegraphics[width=.48\linewidth]{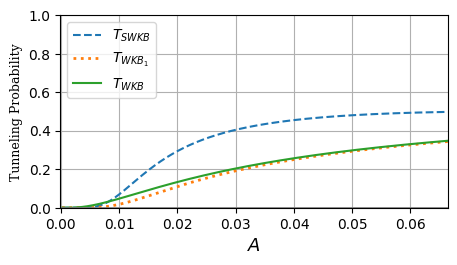}}\hfill
\subfloat[Tunnelling probability $T(A,B,\alpha)$ versus $B$, with $A=0.01$ and $\alpha=1$;]{\label{b}\includegraphics[width=.48\linewidth]{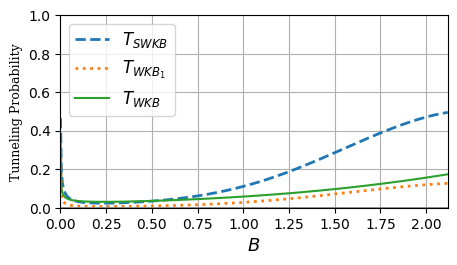}}\par 
\subfloat[Tunnelling probability $T(A,B,\alpha)$ versus $\alpha$, with $A=0.05$ and $B=0.8$.]{\label{c}\includegraphics[width=.48\linewidth]{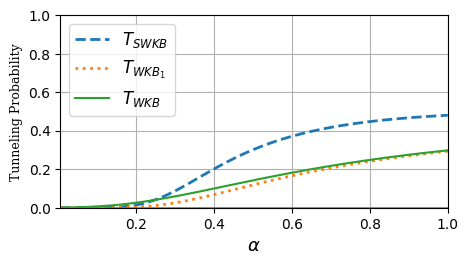}}
\caption{Tunnelling probabilities $T(A,B,\alpha)$ versus \textbf{(a)} $A$, with $B=0.8$ and $\alpha=1$; \textbf{(b)} $B$, with $A=0.01$ and $\alpha=1$; \textbf{(c)} $\alpha$, with $A=0.05$ and $B=0.8$. Blue: SWKB (Picard superpotential); orange: WKB for $V_1(a)$ (SQM potential); green: WKB for the original Chaplygin potential $V(a)$. Turning points were found numerically. The parameter ranges were chosen to ensure the existence of a physical potential barrier. In panels (a) and (b), the plots terminate at the point where the peak of $W_1(a)$ drops to zero, beyond which the barrier vanishes. Panel (c) terminates at $\alpha=1$ to reflect the fundamental constraint $0<\alpha \leq 1$. We notice in all three plots that the tunnelling probabilities obtained by the different approaches are similar for small parameter values and begin to differ for large values.}
\label{figtun}
\end{figure}

The tunnelling probabilities will be functions of $A$, $B$ and $\alpha$ ($T=T(A,B,\alpha)$). We plot them by varying one parameter while holding the other two fixed, as shown in \cref{figtun}. In the first two cases, both $A$ and $B$ go from zero to the value at which $K_{SWKB}$ vanishes, meaning that the peak of the superpotential's barrier corresponds to its roots ($W(a_1)=W(a_2)=0$), so the tunnelling probability goes to $1/2$, as can be seen by equation \eqref{tunprob}.

As noted in \cref{figtun} caption, these ranges were chosen to guarantee a physical barrier exists for all three parameters.

Across all three plots, the probability calculations (SWKB, WKB, and WKB$_1$) somewhat agree for small values of $A$, $B$ and $\alpha$. As the parameters increase (while the others remain fixed), the SWKB and ${WKB}_1$ begin to disagree, with the SWKB probabilities becoming systematically larger. Crucially, throughout this range, the WKB and WKB$_1$ curves remain close to one another, confirming that $V_1(a)$ is a reliable approximation to $V(a)$. The observed differences between SWKB and WKB$_1$ therefore reflect a true discrepancy between the two semiclassical methods, rather than an artefact of the Picard approximation.

The plots show that $T$ increases as each parameter increases, reflecting a shrinking barrier. The exception is $B \to 0$, where $T \to 1/2$, but this limit violates the validity condition \eqref{expansion_chap} and lies outside the regime of interest.

Naturally, the tunnelling probability is maximised for parameter values that minimise both the height and width of the potential barrier. Consequently, quantum creation is probabilistically favoured for universes governed by Chaplygin gas parameters that yield a minimal barrier. This theoretical preference aligns well with current observational paradigms, which strongly favour a spatially flat universe ($k=0$) \cite{aghanimPlanck2018Results2020a,collaborationDarkEnergySurvey2023}; in the minisuperspace framework, a flat geometry eliminates the curvature term entirely, resulting in a barrier-free potential.

\subsection{Domain of Validity}

\begin{figure}
    \centering
    \includegraphics[width=0.5\linewidth]{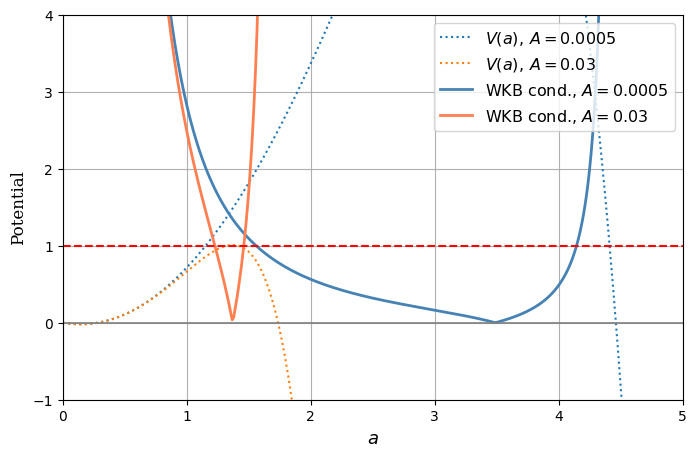}
    \caption{Left-hand side of the WKB validity condition \eqref{wkb_validity} as a function of $a$, for $A=0.0003$ (blue) and $A=0.03$ (orange), with $B=0.8$ and $\alpha=1$. The dashed red line indicates the threshold value of unity, above which the WKB approximation breaks down. We observe that the divergence near the turning points is significantly more pronounced for larger values of $A$.}
    \label{val_cond}
\end{figure}

It is relevant to consider the validity condition for the WKB approximation, which, in the chosen units ($2G=3\pi$), is approximately given by \cite{Bouhmadi_L_pez_2005}
\begin{equation}\label{wkb_validity}
    {\frac{|V'(a)|}{|V(a)|^{3/2}}} \ll 1.
\end{equation}
In \cref{val_cond} we plot the left-hand side of \eqref{wkb_validity} for the values $A=0.0003$ and $A=0.03$, with $B=0.8$ and $\alpha=1$ in both cases, consistent with the parameter values used in \cref{figtun}. The condition approaches or exceeds unity near the turning points in both cases, reflecting the well-known breakdown of the WKB approximation in their vicinity. For larger $A$, this failure is more severe across a wider region of the barrier, directly explaining the growing discrepancy between the WKB and SWKB probabilities observed in \cref{figtun}. Historically, in regimes where \eqref{wkb_validity} is not satisfied, the SWKB approximation has been proposed as a more reliable estimate for shape-invariant systems. However, because our minisuperspace potential lacks this property, we introduce numerical benchmarking in the next subsection.

To rigorously establish the domain of validity for the analytical Picard approximation ($V_1(a)$) and to quantify the physical divergence of the SWKB method, we evaluate two comparative metrics. 

First, the absolute difference, $|\Delta T|$, provides the direct deviation in computed transmission probabilities. Second, we compute the logarithmic difference, $|\Delta \ln T|$. Because the transmission probability was calculated via the formula $T = (1 + \exp(2K))^{-1}$, it asymptotically scales as $T \sim \exp(-2K)$ in the deep-tunnelling limit ($K \gg 1$). Consequently, the logarithmic difference serves as a necessary measure to compare both potentials in the deep-tunnelling region, which are otherwise mathematically masked by infinitesimally small absolute probabilities.

We evaluate these metrics at the extremes of the barrier geometry. Let $x = A, B, \alpha$ denote the varying parameter, and $x_{\max}$ the critical value at which the barrier of the superpotential $W_1(a)$ is fully suppressed and $T_{SWKB}=1/2$. We define the deep-tunnelling regime (massive barriers) as the interval $[0, 0.1 x_{\max}]$, and the suppressed barrier regime as $[0.5 x_{\max}, x_{\max}]$. These intervals were chosen by qualitatively analysing the plots in \cref{figtun}.

Comparing the direct WKB result ($T_{WKB}$) to the one obtained from the Picard approximation ($T_{{WKB}_1}$) isolates the mathematical error introduced by the approximated potential $V_1(a)$. Subsequently, evaluating the SWKB prediction ($T_{SWKB}$) against $T_{{WKB}_1}$ extracts the genuine physical modification introduced by the supersymmetric methodology. These bounds and comparisons are summarised in \cref{tab1}.

\begin{table}[t]
\centering
\small
\renewcommand{\arraystretch}{1.3}
\setlength{\tabcolsep}{5pt}
\begin{tabular}{l l c c c c}
\toprule
\textbf{Varying} & \textbf{Model} & \multicolumn{2}{c}{\textbf{Deep-Tunnelling}} & \multicolumn{2}{c}{\textbf{Suppressed Barrier}} \\
\cmidrule(lr){3-4} \cmidrule(lr){5-6}
\textbf{Parameter} & \textbf{Comparison} & $|\Delta T|_{\max}$ & $|\Delta \ln T|_{\max}$ & $|\Delta T|_{\max}$ & $|\Delta \ln T|_{\max}$ \\
\midrule
$A$ ($B=0.8$, $\alpha=1$)  & WKB vs WKB$_1$  & $1.8 \times 10^{-2}$ & $154.2$ & $9.1 \times 10^{-3}$ & $0.04$ \\
                           & SWKB vs WKB$_1$ & $9.7 \times 10^{-3}$ & $2.67$  & $0.21$               & $0.67$ \\
\midrule
$B$ ($A=0.01$, $\alpha=1$) & WKB vs WKB$_1$  & $3.3 \times 10^{-2}$ & $1.58$  & $4.7 \times 10^{-2}$ & $0.66$ \\
                           & SWKB vs WKB$_1$ & $1.3 \times 10^{-1}$ & $1.32$  & $3.7 \times 10^{-1}$ & $1.39$ \\
\midrule
$\alpha$ ($A=0.05$, $B=0.8$)& WKB vs WKB$_1$  & $6.7 \times 10^{-3}$ & $12.4$  & $2.3 \times 10^{-2}$ & $0.18$ \\
                           & SWKB vs WKB$_1$ & $1.7 \times 10^{-4}$ & $1.58$  & $0.21$               & $0.93$ \\
\bottomrule
\end{tabular}
\caption{Maximum absolute difference ($|\Delta T|_\text{max}$) and logarithmic difference ($|\Delta \ln T|_\text{max}$) across the parameter space. The WKB \textit{vs} WKB$_1$ rows demonstrate the validity of the Picard approximation, which fails in the deep-tunnelling limit but becomes highly accurate for suppressed barriers (as evidenced by the consistent drop in logarithmic error across all parameters).  This statement is supported by the higher-order Picard calculations provided in \ref{3rd_Picard}. In this valid suppressed-barrier regime, the SWKB vs WKB$_1$ rows show a significant physical divergence, confirming the SWKB method predicts higher tunnelling probabilities.}
\label{tab1}
\end{table}

\subsection{Numerical comparison}\label{sec:num_tun}

Because the generalised Chaplygin potential is not shape-invariant, we cannot \textit{a priori} assume that the SWKB transmission probabilities are closer to the exact quantum mechanical values than those obtained via standard WKB. Since this complex potential \eqref{chapapprox} does not admit an exact analytical solution for the tunnelling probability, an independent numerical benchmark is required to arbitrate between the two semiclassical approximations. Adapting the method outlined in \cite{rundquistDirectlyIntegratingSchroedinger2011}, we can directly integrate the Wheeler-DeWitt equation to obtain the tunnelling rates. While this method formally requires the potential to be flat outside the barrier region, a condition not strictly met asymptotically by our potential, it nonetheless converges effectively if the boundary evaluation points are chosen appropriately.

Unlike standard forward-shooting methods, this approach imposes outgoing plane-wave boundary conditions \textit{after} the barrier and integrates backward through the classically forbidden region. This integration is implemented in Python via the \texttt{SciPy} library,  using an explicit eighth-order Dormand-Prince Runge-Kutta method (DOP853) with adaptive step size. We then compare the ratios of the amplitudes of the incoming and outgoing waves to calculate the transmission probability. This backward integration significantly improves numerical stability and circumvents the need to iteratively guess the highly suppressed transmission rate.

Before proceeding, it is worth stating precisely what the quantities computed in the subsequent boundary equations represent, given the constrained, timeless nature of the Wheeler--DeWitt equation. The conserved quantity underlying the transmission rate calculation,
\begin{equation}
J \equiv i\left(\Psi \Psi^{*\prime} - \Psi^* \Psi^\prime\right) 
\end{equation}
where $\Psi'$ denotes the derivative with respect to $a$. It is important to notice that $J$ is a Wronskian-type current defined along the configuration variable $a$ -- a direct consequence of $\Psi^{\prime\prime}=V(a)\Psi$ having a real potential -- and not a probability flux integrated over an external time parameter, since none is available in minisuperspace. Accordingly, $T$ is interpreted as a semiclassical relative weight between the two classically allowed regions of a single, fixed geometry (dust-dominated, pre-barrier, versus accelerating, post-barrier), in the sense of the Vilenkin boundary condition, rather than a transmission probability for repeated scattering events in time.

Furthermore, there is no free ``incident energy'' in this problem analogous to that of ordinary potential scattering. 
In a classically allowed region, the local wave number entering the plane-wave boundary conditions is fixed pointwise by the potential itself:
\begin{equation}
k(a)=\sqrt{-V(a)}\,,\qquad k_\text{well}=\sqrt{-V(a_\text{well})}\,,\qquad k_\text{out}=\sqrt{-V(a_\text{out})}\,.
\end{equation}
The geometry -- encoded entirely in $V(a; A,B,\alpha)$ -- thus determines both the barrier and the wave numbers on either side of it. When $T(A,B,\alpha)$ is evaluated across different parameter values (e.g., in \cref{figtun}), it is the shape of the barrier itself that is being deformed. The wave numbers $k_\text{well}$ and $k_\text{out}$ are recomputed self-consistently from the potential at each parameter value, rather than being held fixed while an independent incident energy is scanned.

\begin{figure}[t]
    \centering
    \includegraphics[width=0.5\linewidth]{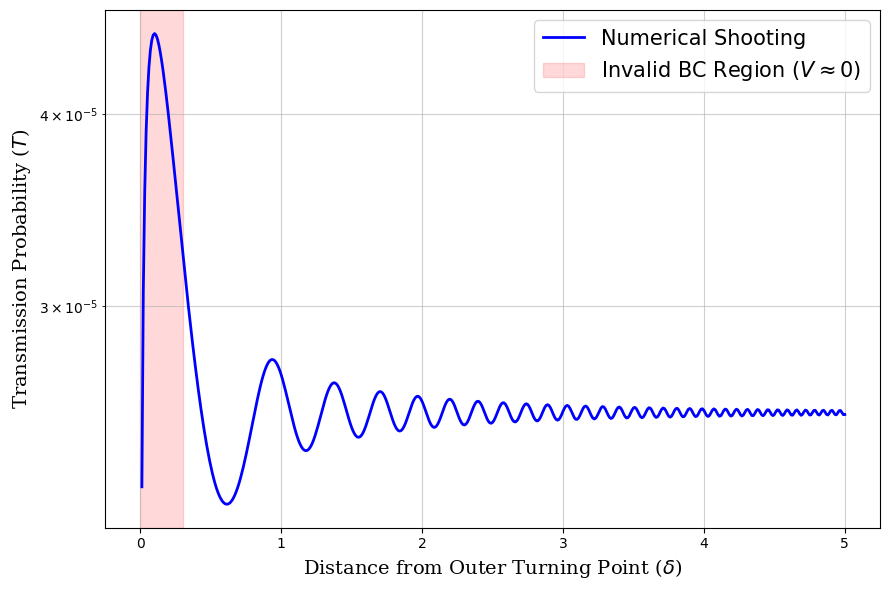}
    \caption{Transmission probability $T$ as a function of the evaluation distance $\delta$ from the outer turning point $a_2$, calculated for $A=0.001$, $B=1$, and $\alpha=1$. The plot demonstrates the numerical stability of the backward integration method. Although the outgoing wavefunction is not a pure plane wave immediately outside the classically forbidden region, the calculated transmission rate exhibits damped oscillations that rapidly converge to a constant value for sufficiently large $\delta$. The red shaded area represents the distances at which our plane-wave boundary condition is not valid. This convergence validates the use of the asymptotic plane-wave boundary condition in our numerical benchmark.}
    \label{fig:numdelta}
\end{figure}
Furthermore, in our numerical implementation, our boundary conditions away from the barrier are
\begin{align}
        \Psi(a_\text{out}) & =Fe^{i k_\text{out} a_\text{out}} \notag \\ 
        \Psi'(a_\text{out}) & =i k_\text{out} F e^{i k_\text{out} a_\text{out}}.
\end{align}
Because the underlying wave equation is linear, the final transmission probability depends solely on the ratio of the amplitudes. Therefore, without loss of generality, we set $F=1$ for simplicity. We then integrate backwards to obtain the solution before the barrier. Approximating the wavefunction as a superposition of incoming and reflected plane waves inside the initial well, we have
\begin{align}
    \Psi(a_\text{well})&=C e^{i k_\text{well} a_\text{well}} + D e^{-i k_\text{well} a_\text{well}} \notag \\
    \Psi'(a_\text{well})&=i k_\text{well} \left(C e^{i k_\text{well} a_\text{well}} - D e^{-i k_\text{well} a_\text{well}}\right). 
\end{align}
Dividing the second equation by $i k_\text{well}$ and summing both equations, we isolate the constant $C$ to obtain the incident amplitude
\begin{equation}
    C=\frac{e^{-i k_\text{well} a_\text{well}}}{2} \left(\Psi(a_\text{well})+\frac{\Psi'(a_\text{well})}{i k_\text{well}}\right),
\end{equation}
and the transmission rate is subsequently given by
\begin{equation}
    T=\frac{k_\text{out}}{k_\text{well}}\frac{|F|^2}{|C|^2}=\frac{k_\text{out}}{k_\text{well}}\frac{4}{\left|\Psi(a_\text{well})-i\Psi'(a_\text{well})/k_\text{well}\right|^2}.
\end{equation}

\begin{figure}[t]
    \centering
    \includegraphics[width=0.55\linewidth]{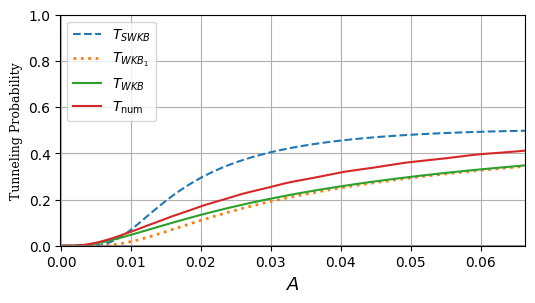}
    \caption{Comparison of transmission probabilities for small parameter variations. For small values of $A$, the numerical method is in close agreement with the WKB result, as expected within the valid semiclassical regime. As $A$ increases and the barrier becomes less prominent, standard WKB begins to underestimate the probability compared to the direct numerical integration. Conversely, in this small barrier regime, the SWKB systematically yields the highest transmission probabilities. Because the numerical method directly integrates the wave equation through the barrier, it successfully bridges the gap between the WKB underestimation and the SWKB over-correction, explained by this potential not being shape-invariant.}
    \label{fig:tun_num}
\end{figure}

Because the potential is not perfectly flat at the evaluation boundaries, the plane-wave ansatz introduces a localised approximation. To minimise this error, we evaluate the inner boundary, $a_\text{well}$, precisely at the minimum of the potential well, where $V'(a_\text{well}) = 0$ and the potential is locally flat. For the outer boundary, we evaluate at $a_\text{out} = a_2+\delta$, where $a_2$ is the outer classical turning point. As demonstrated in our prior analysis of the WKB validity condition \eqref{wkb_validity}, the outgoing plane-wave boundary condition becomes strictly valid in the asymptotic limit $a\to +\infty$. Consequently, although the solution is not a pure plane wave immediately after the barrier, the numerical method successfully converges to a stable transmission rate provided $\delta$ is sufficiently large, as shown in \cref{fig:numdelta}.

We note that while this numerical method relies on plane-wave approximations at the boundaries ($a_\text{well}$ and $a_\text{out}$), sharing similar asymptotic requirements to WKB, it directly integrates the exact wave equation through the barrier. Consequently, it avoids WKB breakdown at the classical turning points, granting it a wider overall domain of validity. We observe, as predicted, that the WKB is valid in the large barrier (small $A$) regime, where the numerical benchmark closely follows this semiclassical result.

As seen in \cref{fig:tun_num}, outside the semiclassical regime, the transmission probabilities follow the hierarchy $T_{WKB} < T_\text{num} < T_{SWKB}$. Standard WKB typically underestimates transmission because it fails to properly account for quantum wave penetration near the classical turning points. The SWKB method attempts to correct this by replacing the standard momentum with the superpotential $W(a)$, which formally regularizes the turning-point singularity. If the potential were shape-invariant, this correction would perfectly capture the sub-barrier dynamics. However, because the Chaplygin potential lacks shape-invariance, the SWKB integral is no longer exact. As our numerical benchmark reveals, for this specific barrier geometry, the supersymmetric formulation overcorrects the WKB deficit, yielding a higher probability. Therefore, our backward numerical integration serves as a necessary benchmark, successfully capturing the exact transmission rate bounded between these two semiclassical methods.

\section{Discussion, Conclusion and Future Prospects}\label{conclusion}

The present work contributes to the application of the SWKB approximation and, more generally, to SQM methods in quantum cosmology. Concretely, we introduced an analytic Picard approximation to obtain the superpotential and subsequently used it to compute SWKB tunnelling probabilities for transitions between a universe in constant expansion and an accelerated one. Comparing these semiclassical results with our numerical integration highlights parameter regimes where the approximations nearly coincide and regions where they differ markedly. Specifically, we established the hierarchy $T_{WKB} < T_\text{num} < T_{SWKB}$, for parameter values outside the domain of validity of the WKB. This demonstrates that while standard WKB underestimates the tunnelling probability due to the turning points divergence, the SWKB method overcompensates on account of the non-shape-invariant nature of the potential. Consequently, the backward numerical integration method proved essential in capturing the true sub-barrier dynamics.

Addressing the fundamental physical relevance of this comparison, our results demonstrate that applying SWKB to quantum cosmology yields far more than a formal mathematical analogy. By imposing the algebraic conditions of supersymmetric quantum mechanics, we introduce non-trivial structural restrictions on the permissible minisuperspace potentials. The most crucial physical consequence is the emergence of a structural bound on the potential's growth that fundamentally alters the late-time dynamics of the universe. Unlike the standard generalised Chaplygin gas model -- which typically transitions from a dust phase, through a tunnelling regime, into an unbounded accelerated expansion (potentially leading to a Big-Rip singularity \cite{Caldwell_2003,Kamenshchik_2013,Bouhmadi-Lopez_bigripescape}) -- the SQM-induced potential dictates a distinctly different cosmological timeline. Specifically, it predicts an initial dust phase, followed by quantum tunnelling, leading into a phase of \textit{transient} accelerated expansion (where dark energy is not permanent), a subsequent return to an effective dust-like phase, and ultimately, a structural halt to unbounded acceleration, allowing the universe to expand without diverging into a permanent dark-energy phase.

Within the minisuperspace framework, this analysis reveals how foundational algebraic symmetries dictate the macroscopic fate of the universe. Because standard WKB systematically underestimates the tunnelling probability, incorporating SQM demonstrates that a transition into this transient dark-energy phase is more viable than previously assumed. Consequently, enforcing rigorous algebraic constraints on the early-universe potential yields profound departures from standard late-time cosmology.

Having established the physical impact of this supersymmetric framework, an intriguing related question is how it might be further generalised: if one could derive other physically motivated superpotentials $W(a)$, their large-$a$ behaviour would directly determine the late-time cosmology.
In practice, this suggests two simple possibilities: polynomial-type $W(a)$ typically give $V(a)\sim W^2\to+\infty$ (a confining wall), while rational or decaying $W(a)$ can make $V(a)$ approach a finite constant or zero and so soften or halt acceleration. If such a $W(a)$ existed, the Chaplygin-driven unbounded acceleration might be softened or avoided, with potential implications for Big-Rip (or related) scenarios \cite{Caldwell_2003,Kamenshchik_2013,Bouhmadi-Lopez_bigripescape,albarranClassicalQuantumCosmology2016, borislavovvasilevClassicalQuantumFate2019}. A natural next step is therefore to evaluate the SQM potential, its classical cosmological implications, and to see whether SQM potentials can change late-time dynamics. This is a promising direction for follow-up work.

Two fundamental steps are required: 1. Establishing an unambiguous correspondence between the effective potentials constructed and the universe's scale factor. 2. Extracting a parametrisation that enables comparison with observational data, such as supernovae (SN), baryon acoustic oscillations (BAO), and other dark energy indicators. This would determine a dark energy that decays after the acceleration phase, as dictated by the new potential. Concretely, dark energy would then depend on the scale factor, exhibiting an inflexion point beyond which it decays. This means proposing a functional form for the dark energy density, derived from a modified potential incorporating supersymmetric ingredients, and comparing it with observational data. 

We should consider cosmological observables that are sensitive to the expansion history, namely:
\begin{enumerate}
    \item \textbf{Type Ia Supernovae (SNe Ia):} luminosity distances constraining $H(z)$ and the deceleration parameter $q(z)$;
    \item \textbf{Baryon Acoustic Oscillations (BAO):} constraints on the comoving angular diameter distance and $H(z)$ at multiple redshifts;
    \item \textbf{Cosmic Chronometers:} differential-age method giving direct measurements of $H(z)$;
    \item \textbf{Integrated Sachs-Wolfe (ISW):} CMB-large scale structure cross-correlations sensitive to evolving gravitational potentials;
    \item \textbf{Growth rate of structure:} redshift-space distortions probing $f\sigma_8(z)$ and the effective equation of state $w(z)$.
\end{enumerate}
If dark energy varies over time, decreasing with cosmic time, then the late-time acceleration should slow down, potentially leading to deceleration or even oscillatory dynamics. Crucially, this transient cosmic acceleration is not merely a formal theoretical curiosity; it constitutes a directly testable physical prediction that is particularly timely, as recent observational data increasingly hint at dynamical, non-permanent dark energy \cite{adameDESI2024VI2025}. Within $\Lambda$CDM (or pragmatic extensions), we may use a combination of the above observables to predict and verify that the acceleration diminishes in accordance with the potential (the orange curve in \cref{fig:3plots1}). A practical route is to employ parametric fits to the evolving equation of state $w(z)$, such as the Chevallier-Polarski-Linder (CPL) ansatz, 
\begin{equation}
    w(a)=w_0+w_a(1-a),
\end{equation}
widely used in dynamical dark-energy analyses \cite{ChevallierPolarski2001,Linder2003,Scherrer2015,LindenVirey2008}. Complementarily, one can adopt \textit{non-parametric} reconstruction techniques (e.g., principal components or Gaussian processes) by combining BAO, SNe Ia, and CMB datasets to test deviations from a constant cosmological term \cite{HutererStarkman2003,XiaLiZhang2013}. Recent works also explore decaying-vacuum $\Lambda(t)$ scenarios constrained by BAO+SNe+CMB+CC \cite{YangWangDai2025}, and quintessence dynamics with multi-probe data \cite{DubeyEtAl2025}.

\ack

DG is grateful to JM and PVM for their guidance and encouragement throughout the research. We thank Mariam Bouhmadi-López for fruitful conversations during the course of our research. DG acknowledges financial support from the Calouste Gulbenkian Foundation, Programme Novos Talentos 2024/2025. JM and PVM acknowledge the FCT grant UID-B-MAT/00212/2020 at CMA-UBI, as well as the COST Actions CA23130 (Bridging high and low energies in search of quantum gravity - BridgeQG) and CA23115 (Relativistic Quantum Information - RQI).

\appendix

\section{From the covariant action to the minisuperspace Lagrangian}\label{app:action_wdw}

For completeness, and to make the origin of equation \eqref{eq:larangian1} fully explicit, we recall its derivation from the Einstein--Hilbert action supplemented by a perfect-fluid matter action of the Hawking--Ellis type \cite{Hawking_Ellis_1973}. Following Mansouri and Nasseri \cite{MansouriNasseri1999}, who carry out this construction for an arbitrary number of spatial dimensions $D$ and arbitrary curvature index $k$, the combined gravity-plus-fluid Lagrangian for the FRW metric $ds^2=-N^2(t)\,dt^2+a^2(t)\,d\Sigma_k^2$ reads, after eliminating a total time derivative,
\begin{equation}
L_D = -\,\frac{V_D}{2\kappa}\left(\frac{a}{a_0}\right)^{D}
\frac{D(D-1)}{N}\left[\left(\frac{\dot a}{a}\right)^2-\frac{N^2k}{a^2}\right]
-\rho\,N\,V_D\left(\frac{a}{a_0}\right)^{D},
\label{eq:LD_app}
\end{equation}
where $\kappa=8\pi G$ and $V_D$ is the volume of the unit $D$-sphere, $V_D=2\pi^{(D+1)/2}/\Gamma\!\big(\tfrac{D+1}{2}\big)$ [cf.\ equations (8)--(19) and (44)--(63) of reference \cite{MansouriNasseri1999} for the full derivation, including the Hawking--Ellis perfect-fluid action $S_M=-\int d^{D+1}x\sqrt{-g}\,\rho$ that produces the last term of equation \eqref{eq:LD_app}].

Specialising equation \eqref{eq:LD_app} to the physically relevant case of a closed ($k=+1$), three-dimensional ($D=3$, $a_0\equiv1$) FRW universe, and using $D(D-1)=6$ and $V_3=2\pi^2$, we obtain
\begin{equation}
\frac{3V_3}{\kappa}=\frac{6\pi^2}{8\pi G}=\frac{3\pi}{4G},
\qquad
\rho\,V_3=2\pi^2\rho ,
\end{equation}
so that equation \eqref{eq:LD_app} reduces \emph{exactly} to
\begin{equation}
L=-\frac{3\pi}{4G}\left(\frac{\dot a^2 a}{N}-Na\right)-2\pi^2 a^3 N\rho ,
\end{equation}
which coincides identically with equation \eqref{eq:larangian1} of our minisuperspace model. The canonical momentum conjugate to $a$, $\pi_a=\partial L/\partial\dot a=-\tfrac{3\pi}{2G}\,a\dot a/N$, together with the vanishing of the coefficient of the (arbitrary) lapse $N$ in the Legendre transform $H=\pi_a\dot a-L=N\mathcal{H}_\perp$, gives the Hamiltonian constraint
\begin{equation}
\mathcal{H}_\perp\equiv\frac{G}{3\pi}\pi_a^2+\frac{3\pi}{4G}\,V(a)=0,
\qquad
V(a)=a^2-\bar\rho(a)\,a^4 ,
\end{equation}
with $\bar\rho(a)\equiv\tfrac{8\pi G}{3}\rho(a)$, yielding equation (3) of the main text.

\subsection*{The Wheeler--DeWitt equation versus the Schr\"odinger equation}

Upon canonical quantisation, $\pi_a\to -i\,d/da$ (we adopt the trivial factor ordering, immaterial at the semiclassical level used throughout this paper), the Hamiltonian constraint becomes the Wheeler--DeWitt (WDW) equation
\begin{equation}
\left[-\frac{G}{3\pi}\frac{d^2}{da^2}+\frac{3\pi}{4G}V(a)\right]\Psi(a)=0 .
\label{eq:WDW_app}
\end{equation}
Equation \eqref{eq:WDW_app} is formally identical to a one-dimensional, zero-energy, stationary Schr\"odinger equation for a fictitious particle of coordinate $a$ moving in the potential $V(a)$ -- an analogy already drawn explicitly, in the closely related context of the de Sitter minisuperspace, by Mansouri and Nasseri \cite{MansouriNasseri1999}. We emphasise, however, that this analogy is only formal, and that the WDW equation differs from an ordinary Schr\"odinger equation in three essential respects.

First, equation \eqref{eq:WDW_app} contains no time derivative and no external time parameter: $t$ in the classical Lagrangian is merely a coordinate label along a worldline in minisuperspace, removable by any reparametrisation $t\to f(t)$, and it disappears altogether upon quantisation. Consequently, $\Psi(a)$ is not a state that \emph{evolves} in time; it is a function on superspace annihilated by the quantum constraint operator $\hat{\mathcal{H}}_\perp$.

Second, the ``zero energy'' in equation \eqref{eq:WDW_app} does not express conservation of energy under time translations generated by a genuine Hamiltonian; it expresses the fact that $\hat{\mathcal{H}}_\perp$ is a first-class constraint, a direct consequence of the reparametrisation invariance of general relativity, which must annihilate all physical states.

Third, as a consequence of the above, quantities such as the tunnelling amplitude and the WKB/SWKB transmission coefficients computed in this work are \emph{not} literal scattering probabilities for a particle traversing a potential barrier in ordinary time. They are semiclassical probability currents defined on minisuperspace via the Hartle--Hawking or Vilenkin boundary conditions imposed on $\Psi(a)$ \cite{hawkinghartlewfofuniverse,vilenkinQuantumCreationUniverses1984,PhysRevD.33.3560}, exactly as in \cite{Bouhmadi_L_pez_2005,MansouriNasseri1999}.

\section{SUSY Harmonic Oscillator}\label{app:susyho}

The harmonic oscillator is a fundamental system in various areas of physics, serving as a straightforward example to illustrate basic concepts. This also applies to the context of supersymmetric quantum mechanics.

The potential of the harmonic oscillator is given by:
\begin{equation}
    V(x) = \frac{1}{2} m \omega^2 x^2,
\end{equation}
Here, $ m $ denotes the particle mass and $ \omega $ the angular frequency. The superpotential is defined as
\begin{equation}
    W(x) = -\sqrt{\frac{m}{2}} \omega x,
\end{equation}
Substituting this into equation \eqref{supot_eq} yields the supersymmetric potential $V_-(x)$:
\begin{align*}
 V_+(x) &\equiv W^2(x) + \dfrac{\hbar}{\sqrt{2m}} \dfrac{dW(x)}{dx} \\
         &= \dfrac{1}{2}m\omega^2x^2 - \dfrac{1}{2}\hbar\omega \\
         &= V(x) - E_0.
\end{align*}
As expected, the original harmonic oscillator potential is recovered, shifted downward by its ground state energy $ E_0 = \frac{1}{2} \hbar \omega $. The ground state wavefunction is calculated as follows.
\begin{equation}
    \psi_0(x) = N e^{+\frac{\sqrt{2m}}{\hbar} \int^x W(x') dx'} = N e^{-\frac{m\omega}{2\hbar} x^2},
\end{equation}
This expression corresponds to the standard ground state of the harmonic oscillator. It can also be confirmed by substitution that
\begin{equation}
    H_+\psi_0=(H-E_0)\psi_0=0.
\end{equation}
We see that $\psi_0$ represents a zero-energy ground state of the supersymmetric quantum mechanics system, indicating that the system exhibits unbroken supersymmetry.

\section{Supersymmetry in Shape-Invariant Potentials: The Pöschl--Teller System}\label{P-Teller}

To illustrate the exactness of the SWKB method for Shape-Invariant Potentials (SIPs), we briefly consider the Pöschl--Teller system. The supersymmetric partner potentials are generated by the superpotential \cite{Hruska_1997}
\begin{equation}
W(x) = \frac{\hbar}{\sqrt{2m}} \frac{\lambda}{a} \tanh\left(\frac{x}{a}\right).
\end{equation}
The resulting potential $V_-(x) = W^2(x) - W'(x)$ forms a potential well (bounded from below), which consequently admits a discrete bound-state spectrum. For this shape-invariant well, the SWKB quantization condition,
\begin{equation}
\int_{x_1}^{x_2} \sqrt{2m\left(E_n - W^2(x)\right)}\, dx = n\hbar \pi,
\end{equation}
with turning points defined by $W^2(x_{1,2}) = E_n$, yields the exact energy eigenvalues: 
\begin{equation}
E_n = \frac{\hbar^2}{2m a^2} \left[ \lambda^2 - (\lambda - n)^2 \right], \quad n = 0,1,\dots,\lambda-1.
\end{equation}
Notably, the ground state gives $E_0 = 0$, confirming unbroken supersymmetry. Thus, the SWKB condition is exact for the bound states of SIPs.

To study quantum tunnelling, one must consider a potential barrier rather than a well, such as the inverted Pöschl--Teller (or symmetric Eckart) barrier $V(x) \propto +\text{sech}^2(x/a)$. Because a physical barrier scatters particles and lacks normalisable bound states, one cannot construct a standard real-valued superpotential for it directly. However, as demonstrated by Sil \textit{et al.} \cite{SIL1994209}, one can bypass this limitation using analytic continuation ($V \to -V$ and $E \to -E$). This mathematical mapping connects the barrier's standard WKB transmission integral to the shape-invariant well's supersymmetric structure.

Sil \textit{et al.} showed that by integrating the square of the well's superpotential,
\begin{equation}
K_{SWKB} = \frac{1}{\hbar} \int_{-x_1}^{x_1} \sqrt{2m(W^2(x) - \tilde{E})}\,dx,
\end{equation}
and calculating $T_{SWKB} = (1 + \exp(2K_{SWKB}))^{-1}$, one obtains the \textit{exact} analytical transmission probability for the Eckart barrier.

This established result serves as a crucial pedagogical baseline: for shape-invariant systems, the SWKB structural substitution ($V \to W^2$) mathematically perfects the semiclassical transmission rate. In our minisuperspace model, however, the general Chaplygin potential lacks shape-invariance, which explains why this SWKB method is not exact.

\section{Next Orders Picard Approximation}\label{3rd_Picard}

A natural extension of this analysis is to compute higher-order Picard approximations for the superpotential. Making use of the general recursive relation in equation \eqref{eq:pic_norder},
\begin{equation}
    W_{n+1}(a) = W_1(a) - \int_0^a W_n^2(s)\,ds,
\end{equation}
the second-order Picard approximation is given by 
\begin{equation}
    W_{2}(a) = W_1(a) - \int_0^a W_1^2(s)\,ds.
\end{equation}
\begin{figure}[t]
    \centering
    \includegraphics[width=0.5\linewidth]{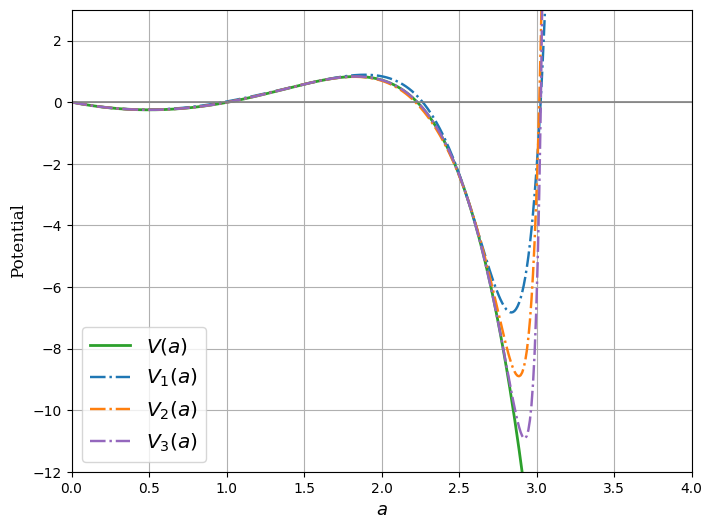}
    \caption{Comparison of the approximated potentials generated by higher-order Picard iterations. The plot shows the exact potential $V(a)$ alongside the 1st (blue), 2nd (orange), and 3rd-order (purple) Picard approximations, $V_1(a)$, $V_2(a)$, and $V_3(a)$, respectively, evaluated at $B=1$, $\alpha=1$, and $A=0.02$. We observe, qualitatively, that the barrier is better approximated at each order.}
    \label{fig:pot_next_ord}
\end{figure}
Substituting the potential 
\begin{equation}
    V(a) = a^2 - aB^\frac{1}{1+\alpha} - \dfrac{1}{1+\alpha}\frac{A}{B^\frac{\alpha}{1+\alpha}}a^{3\alpha+4},
\end{equation} 
and the first-order superpotential 
\begin{equation}
    W_1(a) = \dfrac{a^3}{3} - \dfrac{a^2}{2} B^\frac{1}{1+\alpha} - \dfrac{1}{1+\alpha}\frac{A}{B^\frac{\alpha}{1+\alpha}}\frac{a^{3\alpha+5}}{3\alpha+5},
\end{equation}
into this integral yields:
\begin{align}
    W_2(a) = & -\frac{B^\frac{1}{1+\alpha}a^2}{2} + \frac{a^3}{3} - \frac{B^\frac{2}{1+\alpha}a^5}{20} + \frac{B^\frac{1}{1+\alpha}a^6}{18} - \frac{a^7}{63} - \nonumber \\
    & \frac{AB^\frac{-\alpha}{1+\alpha} a^{3\alpha+5}}{(3\alpha+5)(\alpha+1)} - \frac{A B^\frac{1-\alpha}{1+\alpha} a^{3\alpha+8}}{(3\alpha+8)(3\alpha+5)(\alpha+1)} + \nonumber \\
    & \frac{2 A B^\frac{-\alpha}{1+\alpha}a^{3\alpha+9}}{9(\alpha+3)(3\alpha+5)(\alpha+1)} - \frac{A^2B^\frac{-2\alpha}{1+\alpha} a^{6\alpha+11}}{(6\alpha+11)(3\alpha+5)^2(\alpha+1)^2}.
\end{align}
As is evident, even at second order, the analytical expression for the superpotential becomes quite cumbersome. The corresponding approximated potential is then obtained via the Riccati equation:
\begin{equation}
    V_2(a) = W_2^2(a) + \frac{dW_2}{da}(a).
\end{equation}
Following the same procedure, we can calculate the third-order approximation for the superpotential,
\begin{equation}
    W_{3}(a) = W_1(a) - \int_0^a W_2^2(s)\,ds.
\end{equation}
While its analytical expression is too lengthy to write explicitly, the resulting approximated potentials for these different orders of approximation can be seen plotted in \cref{fig:pot_next_ord}. 

\begin{figure}[t]
    \centering
    \includegraphics[width=0.55\linewidth]{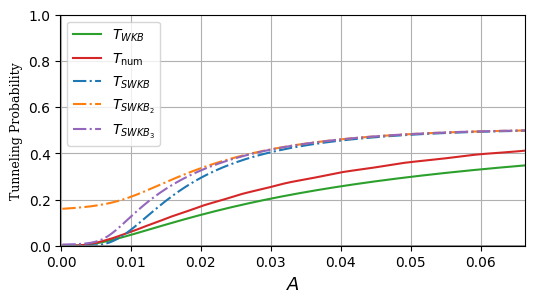}
    \caption{Comparison of the tunnelling probabilities based on the 1st, 2nd, and 3rd-order Picard approximations for the superpotential. For larger values of $A$, where there is a notable discrepancy between standard WKB and SWKB, all Picard orders present similar values. This consistency signals that the discrepancy is a genuine physical feature and not merely an artefact of the approximation. For small values of $A$, the SWKB approximations begin to diverge from one another, reflecting the breakdown of the approximation in this regime. Notably, while all exact probabilities must vanish as $A \to 0$, the second-order SWKB probability plateaus near 0.2. This is because, as seen before, the Picard approximation loses strength in that parameter range}
    \label{fig:tun_next_ord}
\end{figure}

As was the case for the first-order approximation, the calculation of the tunnelling exponent $K_{SWKB_n}$ is performed by integrating the superpotential analytically between the classical turning points, analogous to \eqref{eq:tunswkb_calc}:
\begin{equation}
    K_{SWKB_n} = \int_{a_1}^{a_2} W_n(a)\,da.
\end{equation}
The turning points are again obtained numerically. The tunnelling probabilities are calculated using $T_{SWKB_n}=(1+\exp(2K_{SWKB_n}))^{-1}$. The obtained probabilities for the 1st, 2nd, and 3rd-order Picard approximations are plotted in \cref{fig:tun_next_ord}. We note that in the parameter range where a large discrepancy exists between the WKB and SWKB calculations, the different orders of the Picard approximation remain highly consistent. As previously anticipated, this confirms that the discrepancy is not an approximation error, but rather a genuine physical divergence between the two methods.

\section*{References}

\bibliographystyle{iopart-num}
\bibliography{reference_cqg_new}

\end{document}